\documentclass[namedreferences]{solarphysics}
\usepackage[hyperref,optionalrh,solaromanenum]{spr-sola-addons} 
\usepackage[pdftex]{graphicx}            
\usepackage{amssymb}                     
\usepackage{color}                       
\usepackage{epstopdf}

\usepackage{comment}
\ifx \doiurl  \undefined \def \doiurl#1{\href{http://dx.doi.org/#1}{\url{#1}}}\fi
\ifx \adsurl  \undefined \def \adsurl#1{\href{http://adsabs.harvard.edu/abs/#1}{\textsf{ADS}}}\fi

\chardef\us=`\_

\begin{document}
\begin{article}

\begin{opening}

\title{Reconstruction of the Sunspot Number Source Database and the 1947 Zurich Discontinuity}

\author[addressref={1},corref,email={frederic.clette@oma.be}]
{\inits{F.}\fnm{Fr{\'e}d{\'e}ric}~\lnm{Clette}}
\author[addressref={1}]{\inits{L.}\fnm{Laure}~\lnm{Lef{\`e}vre}}
\author[addressref={1}]{\inits{L.}\fnm{Sabrina}~\lnm{Bechet}}
\author[addressref={2}]{\inits{R.}\fnm{Renzo}~\lnm{Ramelli}}
\author[addressref={3}]{\inits{M.}\fnm{Marco}~\lnm{Cagnotti}}

\runningauthor{Clette et al.}
\runningtitle{Sunspot Number Database and the 1947 Zurich Discontinuity}

\newcounter{affiliations}
\address[id=1]{Royal Observatory of Belgium, 3 Avenue Circulaire, 1180 Brussels, Belgium}
\address[id=2]{Istituto Ricerche Solari Locarno (IRSOL), Universit{\`a} della Svizzera italiana, Via Patocchi 57, 6600 Locarno, Switzerland}
\address[id=3]{Specola Solare Ticinese, Via ai Monti 146, 6605 Locarno, Switzerland}

\begin{abstract}
The recalibration of the sunspot number series, the primary long-term record of the solar cycle, requires the recovery of the entire collection of raw sunspot counts collected by the Zurich Observatory for the production of this index between 1849 and 1980.

Here, we report about the major progresses accomplished recently in the construction of this global digital sunspot number database, and we derive global statistics of all the individual observers and professional observatories who provided sunspot data over more than 130 years.

First, we can announce the full recovery of long-lost source-data tables covering the last 34 years between 1945 and 1979, and we describe the unique information available in those tables. We then also retrace the evolution of the core observing team in Zurich and of the auxiliary stations. In 1947, we find a major disruption in the composition of both the Zurich team and the international network of auxiliary stations. 

This sharp transition is unique in the history of the Zurich Observatory and coincides with the main scale-jump found in the original Zurich sunspot number series, the so-called ``Waldmeier'' jump. This adds key historical evidence explaining why methodological changes introduced progressively in the early 20$\rm ^{th}$ century could play a role precisely at that time. We conclude on the remaining steps needed to fully complete this new sunspot data resource.   
\end{abstract}


\end{opening}
\section{Introduction} \label{S-Intro} 
Our knowledge of the long-term evolution of the solar cycle is largely based on the historical observations of sunspots since the newly invented telescope was aimed at the Sun for the first time in 1610. Two main indices were built from those sunspot observations. The sunspot number (hereafter SN) was initiated by Rudolf Wolf in 1850 \citep*{Wolf1856,Friedli2016}. This daily index combines the total group count and the total spot count and its goes back to 1700. Much more recently, \citet{HoytSchatten1998a,HoytSchatten1998b} introduced the sunspot group number (hereafter GN), which only uses the total group count, but was constructed back to the very first telescopic observations in 1610. Both indices are abundantly used by most studies of the long-term evolution of solar activity and Sun-Earth relations, as constraints for validating physical models of the solar dynamo, and for calibrating various parameters relevant to space weather and space climate (geomagnetic and ionospheric indices, cosmogenic radionucleides).

However, significant disagreements between the sunspot number and group number series over their common time interval indicated that either series or both suffered from inhomogeneities. This prompted various efforts to identify flaws and biases in both series, which led to the release of the first revised versions of the group number \citep[``backbone'' GN]{SvalgaardSchatten2016} and of the sunspot number \citep[SN Version 2.0]{CletteEtal2014,CletteLefevre2016}. Regarding the GN, further corrections and improvements have been proposed over recent years, but we will not develop this ongoing work here \citep[see e.g.][]{ChatzistergosEtal2017,WillamoEtal2017,SvalgaardSchatten2016, Svalgaard2020,UsoskinEtal2021}. However, a key element that supported this revision effort was the expansion and correction of the GN database collecting all original observed group counts \citep{VaqueroEtal2016}. This work, which started from the original database assembled over many years by \citet{HoytSchatten1998a,HoytSchatten1998b}, is still continuing now, and already allowed new improved reconstructions of the GN directly from the base source data. As highlighted by \citet{Munoz2019}, the recovery of all existing historical observations is crucial for future progresses in such reconstructions of past solar activity. 

By contrast, the current revised SN series was reconstructed from source data only for the recent decades, since 1981, when the production of the SN moved from the Zurich Observatory to the Royal Observatory of Belgium, where it is still maintained today \citep{CletteEtal2007,CletteEtal2016}. Indeed, the data processing was then computerized, and all collected data from the worldwide network of contributing stations are preserved in digital form (more than 500,000 observations from 285 stations). On the other hand, for the entire Zurich period before 1981, the corrected SN series was obtained by deriving and applying correction factors to the original Zurich SN series, as provided by Wolf and his successors \citep{CletteEtal2014,CletteLefevre2016}. This approach already allowed to correct the main flaws present in the original SN series and affecting long segments of this series, in particular a sharp 18\% upward jump in 1947 \citep[see][for the details]{CletteLefevre2016}, but it faces limitations for finer corrections.

This more indirect and limited approach was imposed by two main constraints that are specific to the history of the sunspot number. While the GN was directly built from the whole set of available observations, the Zurich SN was mostly based on the sunspot counts from the Zurich Observatory, which acted as pilot station. The data from auxiliary stations were mostly used to fill in the daily gaps due, e.g., to bad weather in Zurich, and they thus only played a secondary role in the production of the early part of the SN \citep{CletteEtal2014,DudokEtal2016,Friedli2016,Friedli2020}. As a consequence, the sources of inhomogeneity are predominantly associated with a single reference station, and are thus very different from the GN, which requires other diagnostics.

However, the other major restriction was the absence of a global digital database of the source data collected by Wolf and his successors. As we will describe later in this article, only part of those data were published, and none of those data were converted into digital form. The inaccessibility of the Zurich source data prevents researchers from getting access to a huge amount of detailed information and to essential metadata. The recovery of this vast collection can feed full statistical analyses by current state-of-the-art methods and lead to an improved index, independent of all assumptions and practices adopted over the years by Wolf and his successors at the Observatory of Zurich. 

This is what motivated a collective effort to recover and digitize all those original source data. Major progresses have been accomplished over the past few years. In this article, we report on those major advances. In Section \ref{S-digi}, we first present the global digitization of the published data, available in printed form, and complemented by deeper archives of hand-written logbooks. Based on the resulting global chronology of all contributing observers assembled in Section \ref{S-Chrono}, we summarize the temporal evolution of the sources on which the SN was founded. In Section \ref{S-WaldTab}, we then present the recent recovery of the long-lost Waldmeier archives, and we describe the contents of those new tables. Based on the now-continuous historical timeline, we show the occurrence of a double discontinuity in the composition of the Zurich team of observers (Section \ref{S-ZuObs}) and the network of auxiliary stations (Section \ref{S-AuxSta}). In Section \ref{S-Conclu}, we finish by concluding on the overall Zurich history emerging from this early exploration of the new SN database, and on the prospects and upcoming tasks.

\section{Complete Digitization of Published Tables (1849-1944)} \label{S-digi}

\subsection{The Zurich Printed Data: Full Survey of the Mitteilungen}

The Zurich sunspot number produced by Wolf and his successors is based on three types of data: 
\begin{itemize}
	\item the raw counts from the Zurich staff: essentially, the director and the assistants in Zurich, and also in the course of the $20^{th}$ century, other assistants stationed in the Arosa and Locarno observatories in southern Switzerland.
	\item the counts sent to the Observatory of Zurich by external auxiliary observers, either individual solar observers or professional observatories.
	\item the historical observations collected by Wolf over the course of his entire career, which extend the first two sets of data before 1849 and back to 1610. Most of those numbers were recounted by Wolf himself from original documents \citep{Friedli2020}.
\end{itemize}
Most of this material was published on a yearly basis in the bulletins of the Zurich Observatory, the Astronomische Mitteilungen der Eidgen{\"o}ssischen Sternwarte Z{\"u}rich (hereafter Mitteilungen). This is a fundamental resource for any future recomputation of the SN series. As noted in the introduction, a large part of those data were never directly used for the production of the sunspot number, as on most days, the SN was simply the raw Wolf number from the Zurich Observatory. 

\begin{figure}
	\centering
	\includegraphics[width=0.8\linewidth]{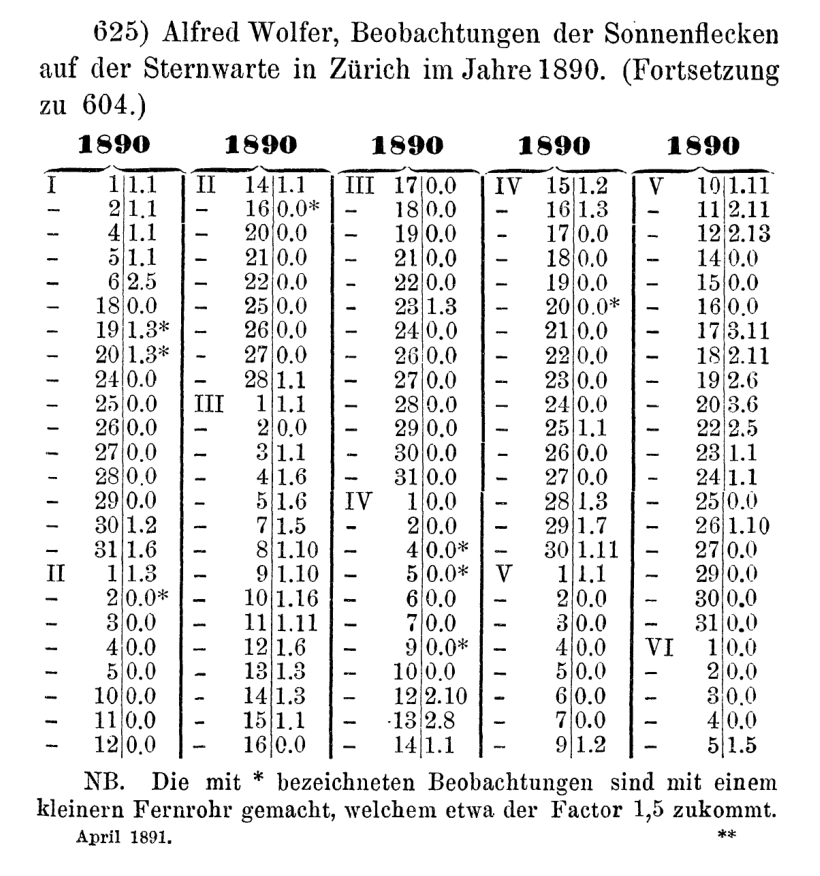}
	\caption{Facsimile of a typical yearly table, as published in the Mitteilungen (first page going up to early June). This table lists all daily observations from A.Wolfer for the year 1890. Each column gives the date followed by the total number of groups and total number of spots, separated by a dot. A star symbol is added for some days, and marks the days when the observations were made occasionally with a different telescope (On these days, Wolfer used a small portable ``Parisian'' telescope with a 40\,mm aperture).}
	\label{fig:MittTable}
\end{figure}

In each issue of the Mitteilungen, the source data are listed in a series of numbered rubrics at the end of the issue. The rubric series starts in 1857 (Volume 3, page 126) and ends in 1930 (Volume 122, page 41), at the 1727$\rm ^{th}$ entry, forming all together a very comprehensive census of all data collected by the Zurich Observatory. Systematic observations by the Zurich observers (with the director and his assistants listed separately from 1870 onward) and by auxiliary observers are presented in yearly tables (Figure \ref{fig:MittTable}) with, for each observed day, the number of groups $g$ and number of spots $s$, in the standard format $g.s$. The table is preceded by a brief description of the observer, mainly his/her name, the general location (city), and in most cases, the kind of telescope used for the observations (aperture, focal length and magnification). Symbols are sometimes added in the table to mark changes on a daily basis. The symbol may identify a specific observer when there are several observers working in the same observatory. In other cases, it marks a change of location or instrument. A prominent example involves Wolf himself, who observed either with the standard 83\,mm Fraunhofer refractor mounted permanently at the Zurich Observatory or with smaller portable refractors \citep{Friedli2016, Friedli2020}. This auxiliary information can thus prove essential for the proper exploitation of the raw data.

Sometimes, when Wolf includes a new observer who already collected spot counts over many years, a long multi-year table is published with all those past observations. Key examples are the tables for Staudacher (Vol. 4, 1857), Schwabe (Vol. 10 , 1859), Flaugergues (Vol. 13 , 1861), Carrington (Vol. 35, 1873) or Pastorff (Vol. 36, 1875). Finally, next to the tables, many rubrics mention small isolated data sets, or even unique spot counts found in old documents during searches that Wolf did in libraries all over Europe. These are mostly single sunspot sightings that are embedded in textual descriptions, e.g. spots noticed at the occasion of solar eclipses. Although they may be individually important, all together, they form only a tiny fraction of Mitteilungen data ($< 1\%$), and they are less exploitable because they cannot be calibrated.

\subsection{The Digitization: First Milestone}
So far, this large collection of data was completely inaccessible in digital form, by contrast with the GN database, which includes all raw group counts collected by \citet{HoytSchatten1998a,HoytSchatten1998b} and was recently expanded by \citet{VaqueroEtal2016}. Although there is a rather wide overlap between the GN and SN data and many observers are common to both data sets, the GN database unfortunately contains only the number of groups. Therefore, the number of spots can only be found in the Zurich data, as it was required to compute the SN. This thus motivates the construction of a complete SN database, equivalent to the existing GN database. 

As a first major step, in 2018, a full encoding of the Mitteilungen data tables was undertaken at the World Data Center Sunspot Index and Long-term Solar Observations (SILSO), with the help of students for the bulk encoding work. By the end of 2019, all the data tables have been digitized, forming the first version of the SN database, which includes all data published between 1849, when R. Wolf undertook the production of the sunspot number, and 1944, when the last director, Max Waldmeier, decided to cease publishing raw data in print. This database now contains 205,000 individual daily sunspot counts. Isolated numbers mentioned in textual rubrics are not yet included, but we plan to add them later on, for the sake of historical completeness.

Next to the daily separate counts of spots and groups, the database includes metadata derived from annotations in the printed tables. When daily symbols indicated regular changes of observers or instruments and when each subset included a large number of days, we split the data included in common tables, and attached the subsets to distinct observers. So, an observer may appear in different incarnations, corresponding to different instruments and/or locations, which thus require a different calibration and should not be mixed. 

Currently, this first major input to the SN database is subjected to a thorough quality control, fixing typos, date inconsistencies and occasional ambiguities in observer names. Meanwhile, we looked for other data sources that can help recovering information that proved to be missing in the Mitteilungen. One of the gaps happens in the early part of the SN database.

\subsection{Wolf's Sourcebook and Wolfer's Global Register}

Indeed, before 1870, the information about the core observations made by Wolf and his assistants is incomplete. A single ``master'' yearly table contains all the counts used to produce the sunspot number. It thus consists mainly of the counts made by Wolf, which are thus largely complete. On the other hand, data from other observers, assistants or external observers, are only inserted on days when the primary observer could not observe. As a consequence, between 1864, when the first assistants were recruited, and 1869, only a small fraction of the data from the Zurich assistants appear in the Mitteilungen, as Wolf's own data fill a majority of days. 

Moreover, before 1864, Wolf's main auxiliary observer was Samuel Heinrich Schwabe. However, although a significant fraction of the daily counts were from Schwabe, Wolf did not mark them in the published tables before 1859, as he first considered Schwabe's numbers fully equivalent to his own. This now makes it impossible to distinguish Wolf's primary counts from rescaled numbers from Schwabe during the first 10 years of the Wolf series. This important information about the primary Zurich observers is thus largely incomplete between 1849 and 1870. 

\begin{figure}
	\centering
	\includegraphics[width=1\linewidth]{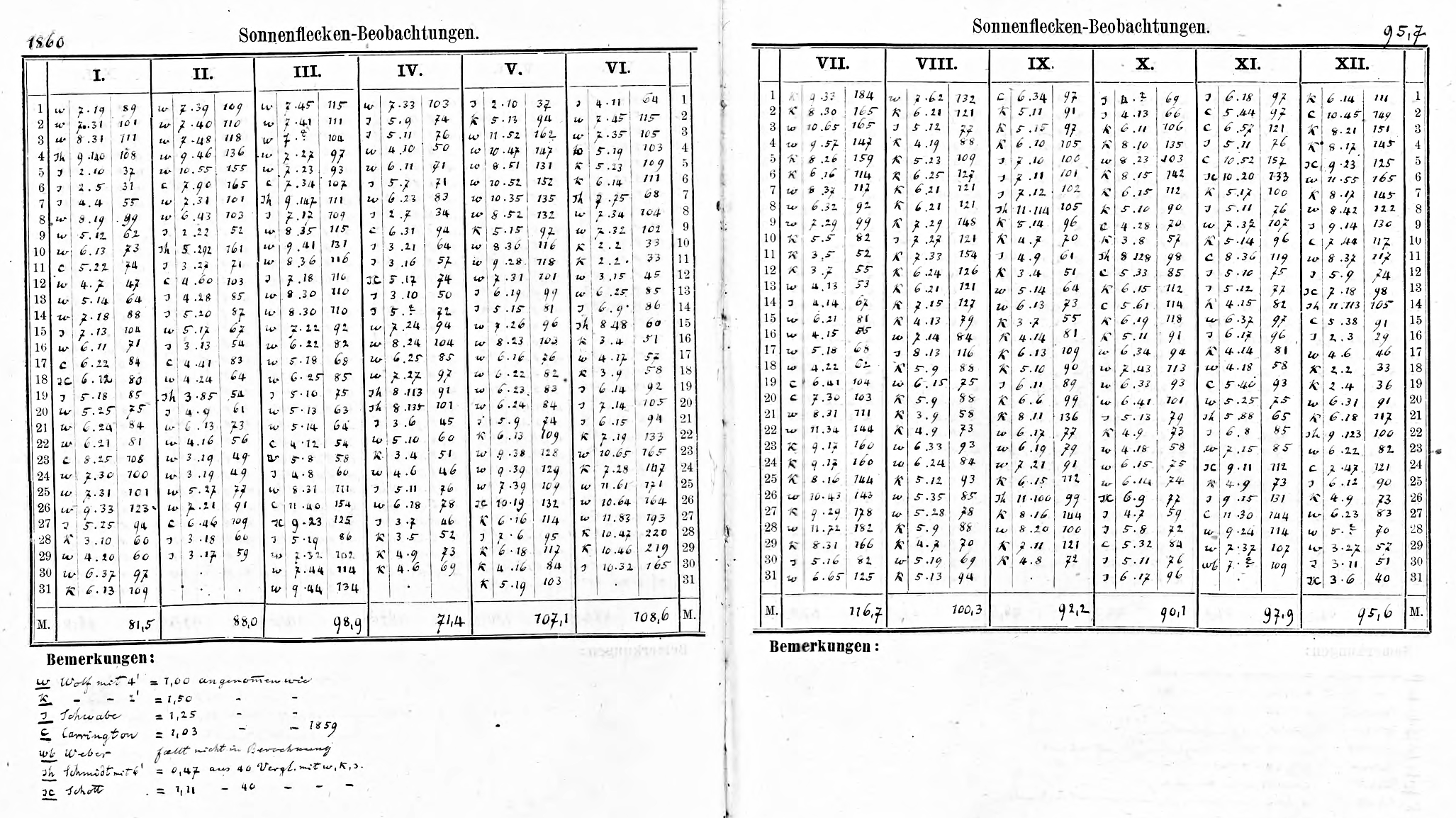}
	\caption{Facsimile of the table for 1860 in Wolf's hand-written sourcebook, which covers the period 1610 to 1877 (ETH  catalogue entry Hs368:46). The layout is similar to the equivalent yearly table published in the Mitteilungen, but it contains very important additional information. Symbols indicate for each day, from which observer the daily sunspot number was obtained. One can see that most of the observations were from Wolf, as primary observer. For each auxiliary observer, including here Schwabe and Carrington, a list at bottom left mentions the personal k coefficient that was used to rescale the raw numbers, to match the scale of Wolf's own numbers.}
	\label{fig:Wsourcebook}
\end{figure}

Fortunately, two additional sources that provide full tables of the base counts were preserved, and are archived at the ETH Zurich University Archives of the Eidgen{\"o}ssische Technische Hochschule (ETH). One of them is the so-called Wolf's sourcebook \citep[catalogue entry Hs368:46]{Wolf1878}. Those handwritten yearly tables gather all daily numbers forming the sunspot number series from 1610 to 1877 (Figure \ref{fig:Wsourcebook}). In fact, these are the master tables assembled by Wolf \citep{Friedli2016}. Those tables provide two kinds of unique information. Firstly, right from the start of Wolf's yearly census in 1849, they include symbols identifying the source observer for each daily number. This additional information will thus allow to remove the ambiguity in the early Mitteilungen tables. Moreover, each yearly table indicates the personal k coefficient that was actually used by Wolf, a precious information that can be crossed with the few occasional mentions by Wolf of changes in his k calculations. 

Moreover, as the copying and typesetting process for the publication in the Mitteilungen most probably led to errors and typos, the original sourcebook provides the ground truth and will allow fixing those occasional mistakes in the master database. Thanks to the efforts of the Wolf Gesellschaft \citep{Friedli2016}, Wolf's sourcebook was digitized from 1849 to 1877, when the collection ends. While the tables can now be consulted online at URL \url{http://www.wolfinstitute.ch/data-tables.html}, this extended information must still be merged with the primary Mitteilungen database. This work is now in preparation. Finally, the yearly tables in the sourcebook actually go back to the very first sunspot observations in the early 17$\rm ^{th}$ century. Although this part is less substantial, those data tables for years before 1849 must still be digitized.

\begin{figure}
	\centering
	\includegraphics[width=1\linewidth]{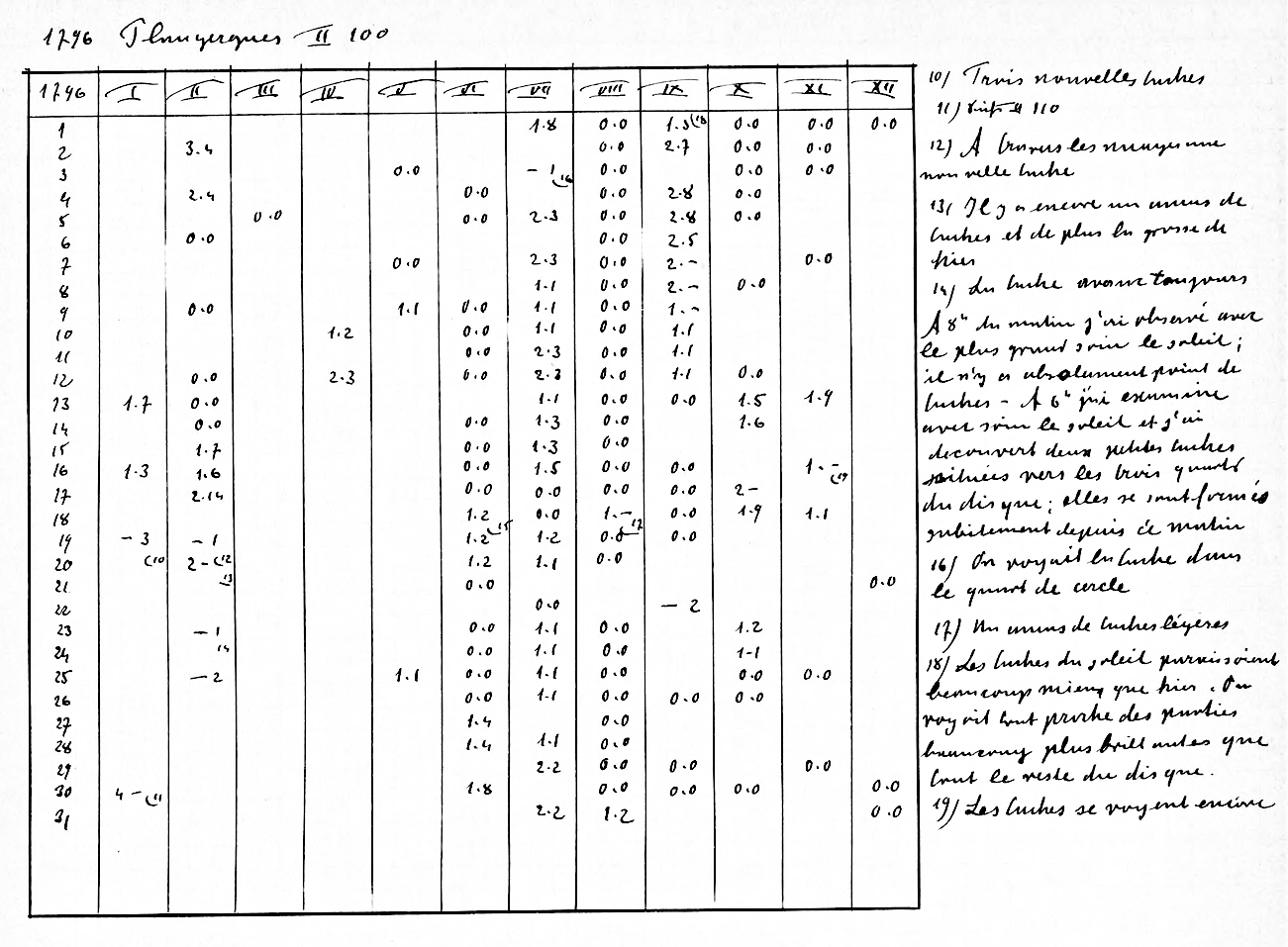}
	\caption{Facsimile of one page from the register of hand-written tables compiled by R. Wolf and continued by A. Wolfer, and covering the entire period 1610 - 1908 (ETH catalogue entry Hs1050:227). This page shows the yearly table for Flaugergues in 1796. The layout is similar to the yearly tables in Wolf's sourcebook, but here, all daily observations are listed for each observer. On the right, literal citations and detailed indications are often included to clarify the interpretation of the tabulated numbers. This series of tables thus gives a complete and well-standardized view of all data collected by Wolf and Wolfer, including data that were not used to produce the daily sunspot number, and also data and metadata that were not published in the Mitteilungen.}
	\label{fig:W-Wtables}
\end{figure}

However, like in the Mitteilungen, the sourcebook does not contain the full set of raw observations collected by Wolf from the auxiliary observers and from his assistants, between 1849 and 1870, in particular, the observations from Schwabe. However, a larger set of handwritten tables also exists at the ETH Zurich University archives \citep[catalogue entry Hs1050:227]{Wolfer1909}. This series is a standardized compilation of all data and metadata published in the Mitteilungen, up to 1908 (Figure \ref{fig:W-Wtables}). This huge register was first produced by Wolf, and after Wolf's death in 1893, it was continued by A. Wolfer and his assistants until 1909, as a base for a global verification of the sunspot number series. In this collection, there is a separate table by observer and by year. Therefore, the full data set is included, even data that were never used for the calculation of the daily Zurich sunspot number. In particular, there are also many data series from before 1700, which were never used by Wolf, as he decided to compute the sunspot number only from 1700 onwards. 

Still, the tables in this complete register may prove invaluable for crossing this information collected long ago by Wolf with other recovered observations of the same observers. They also indicate which data were known by Wolf and his collaborators at the epoch when they produced the Zurich numbers. The scanning of this large set of tables is now planned at the ETH Library in Zurich. When this step will be completed, the encoding into a database will require substantial additional work.

\begin{figure}
	\centering
	\includegraphics[width=1\linewidth]{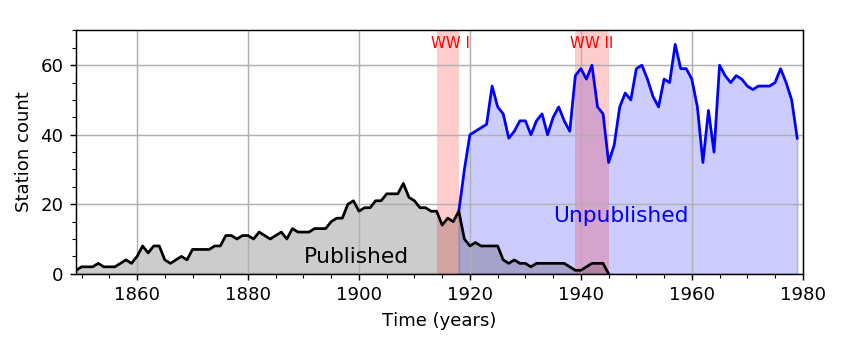}
	\caption{Evolution of the number of stations for each year contained in the data tables published in the Mitteilungen of the Zurich Observatory (gray curve). After 1919, when the Zurich Observatory ceased to publish all the data, the total number of contributing stations is plotted in blue, based on the annual list of stations. Between 1919 and 1944, the data from a subset of observers were still included, but after 1945, none of the source data were published. The two vertical shaded bands mark the two world wars, which both definitely left an imprint on the Zurich sunspot data set.}
	\label{fig:MittStat}
\end{figure}

\section{Chronology of the Data} \label{S-Chrono}
Although series of data and metadata still needs to be added, the database is now largely complete between 1849 and 1944, and we have now already a complete chronology of all the observers who provided data to the Zurich Observatory between 1849 and 1980, i.e. during the entire Zurich era.
This allows us to derive some global statistics of the observers and the time interval over which they were active, which provides very interesting new insights in the construction of the Zurich SN. 

Figure \ref{fig:MittStat} shows the number of stations for each year. This illustrates the evolution of the input data, as published in the Mitteilungen. The number of stations steadily increased from 1865 to 1896, when it reaches about 20 stations and then drops slightly, but remaining above 15. This corresponds to the continuous recruiting of new additional external observers by Wolf and later by Wolfer. This evolution is completely disrupted in 1919. At the end of World War I (WWI), Wolfer adds many new observers. The number of stations passes the 40 mark, doubling the size of what becomes a true international network. However, probably for financial reasons, Wolfer then decides not to publish all data anymore \citep{Friedli2020}. Only the numbers from the Zurich observers and 7 to 9 primary external observers are still published each year. Although some of those privileged external observers had been important long-term contributors by 1919, the selection criteria are unclear and were not explained by Wolfer. 

But another drop of the number of tabulated data happens in 1926, when William Otto Brunner succeeds Wolfer as director of the Zurich Observatory. Brunner then decides to publish only the data from the Zurich team \citep{Brunner1927}. None of the data from the network are published after that year. The only exception is Karl Rapp, a private observer, who observed in Locarno, Switzerland, from 1940 to 1957. Rapp was actually trained in the same way as assistants at the main observatory in Zurich, and was thus treated as an internal observer over his whole observing career. Although Brunner states in 1927 that the external data from auxiliary stations are archived and can be consulted on request \citep[page 188]{Brunner1927} \citep[Section 3.3]{Friedli2020}, searches undertaken over past years failed to recover those archives. So far, only the data for 1944 were found in a single unpublished manuscript, referenced Hs1050:14 in the ETH Zurich University archives, which contains all calculation sheets for that single year \citep{Friedli2020}.

Then in 1945, when Max Waldmeier becomes the new director, the publication of source data ceases completely, as can be seen in Figure \ref{fig:MittStat}. By then, the volume of data collected in Zurich had further increased, with almost 60 contributing stations (blue curve in Figure \ref{fig:TimelinesStations}), making their publication bulky and costly. The Mitteilungen then switch to a different format. The thick yearly volumes become a series of shorter thematic issues, with articles about diverse research topics developed by Waldmeier. The sunspot number gets a more limited space, compared to the earlier volumes published by Wolfer and Brunner, which were almost entirely dedicated to sunspots. Again, during this last period of the Zurich history, all the original data were saved like before in archives at the observatory in Zurich. 

However, since the closing of the Zurich Observatory in 1980, those archives somehow went lost. This created a major 35-year data gap in the raw data collection on which the Zurich sunspot number is based. This wide gap falls at a critical moment, as one of the main scale jumps identified in the Zurich series falls in 1947, thus precisely within this time interval \citep{CletteEtal2014, CletteLefevre2016}. The raw input data are thus essential to reconstruct the methodological changes that took place in Zurich at that epoch and may have caused this inhomogeneity. Moreover, this gap creates a critical missing link between the early Zurich epoch, up to Brunner, and the modern international sunspot number produced in Brussels since 1981, for which all data are preserved in a computer-accessible digital database.

\section{The Original Waldmeier Source Tables (1945-1980)} \label{S-WaldTab}

\subsection{A Serendipitous and Complete Recovery}
Fortunately, in late 2018 and early 2019, a serendipitous finding by the staff of the Specola Solare Ticinese Observatory in Locarno ({\url https://www.specola.ch/e/)}, followed by subsequent searches, allowed to recover the entire Waldmeier data archive (1945\,--\,1979), which was in fact dispersed over three locations: the Specola Observatory (26 years, 1945\,--\,1970), the Royal Observatory of Belgium in Brussels (4 years, 1971\,--\,1974), and in the deep storage of the ETH Zurich University archives in Zurich (5 years, 1975\,--\,1979). This dispersion seems to be due to the rather tumultuous closure of the Zurich Observatory \citep[for an evocation of that transition, see][]{Stenflo2016}. Except for copies of the years 1975\,--\,1979 on microfiches at the ETH archives, the fragmented original collection was also stored without inclusion in any inventory or catalogue. 

This recovery is a breakthrough, and given the amount of data collected over those 35 years, it will keep researchers busy for many years. Indeed, we estimate that those tables contain about 350,000 individual daily numbers, thus more than in all published tables from 1849 to 1944. In a first step, all the elements of this archive were brought together again at the ETH Zurich University archives. They are now fully cataloged \citep{Waldmeier1980}, and the ETH archives have completed the digitization of the whole collection in 2020. The scans of all tables are now accessible online on the digital platform for manuscript material from Swiss libraries and archives at \url{https://www.e-manuscripta.ch/} (ETH catalogue entry Hs 1304.8). Now, in order to make all the data computer-readable, all those tables need to be encoded. This work has now just started at the Royal Observatory of Belgium.   

\begin{figure}
	\centering
	\includegraphics[width=1\linewidth] {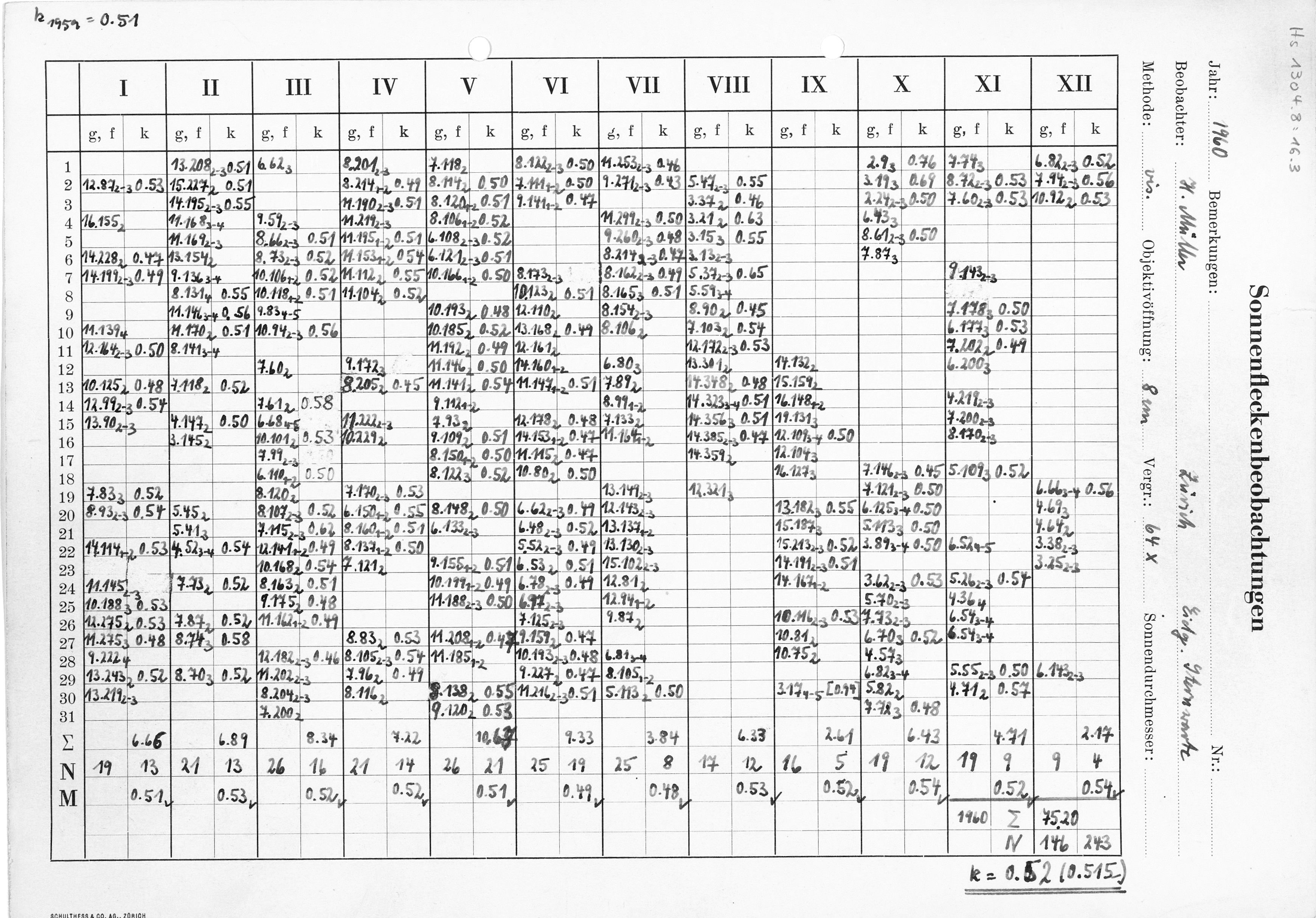}
	\caption{Facsimile of a typical handwritten yearly table from the complete 1945\,--\,1980 collection of source tables that was recovered in 2018\,--\,2019. This table lists the data from H. M{\"u}ller, one of the assistants observing at the Zurich Observatory with the standard 8-cm Fraunhofer refractor, for the year 1960 (ETH catalogue entry Hs1304.8:16.3; DOI: {\it 10.7891/e-manuscripta-87290}). For each day, the table gives the number of groups, the total number of spots, the calculated personal k value relative to the primary observer (Waldmeier), and a sky quality index. For each column, monthly sums and the mean k coefficient are given at the bottom. The yearly totals and the mean k coefficient for the whole year are appended at the lower right. Here, k equals 0.52 and thus differs by more than 15\% from Waldmeier's target value of 0.6, revealing a significant dispersion of the Wolf numbers from the assistants, although they were expected to be closely aligned on the primary observer.}
	\label{fig:MullerTable1960}
\end{figure}

\subsection{The Waldmeier Yearly Tables: a Key to the Zurich Method}
The Waldmeier archive consists in yearly handwritten tables, one per observer, and each one on a separate sheet. Over the period 1945\,--\,1980, there was an average of 50 stations each year. All tables adopt the same standard format, with one column per month. Figure \ref{fig:MullerTable1960} illustrates the typical layout of one sheet, here with the table for H. M{\"u}ller, one of the Zurich observers, for the year 1960. External auxiliary stations are presented with exactly the same layout. Each table lists all daily observations provided by the observer. The number of spots and groups are given separately, exactly like in the tables published earlier in the Mitteilungen.

This essential piece of information, which was so far entirely lost, will allow to determine for each day exactly how the observers were separating sunspot groups, on the one hand, and counting sunspots on the other hand. In particular, it will help clarifying and quantifying the use of weighted sunspot counts, an alternate counting method adopted by the Zurich observers, in particular by Waldmeier himself. This alternate counting rule, in which large spots with extended penumbra are counted as more than 1, is suspected to be the cause of the 18\% upward jump that affected the original SN series in 1947 \citep{CletteEtal2014,CletteLefevre2016, SvalgaardEtal2017}. Indeed, recent double counts, using the regular Wolf formula or weighted counts, were made at the Specola Observatory during several years, between 2003 and 2015, and led exactly to the same inflation of the sunspot number as the one found in the Zurich series after 1947 \citep{CletteEtal2014,SvalgaardEtal2017}. The recovered tables are thus providing the same kind of evidence, but over 35 years, including the epoch when the jump occurred.

The tables also include the monthly and yearly mean k personal coefficients computed by the Zurich Observatory, a very important piece of metadata to understand how Zurich was treating the source observations. In particular, k coefficients are given for all Zurich assistants, and also the associated observers of the Specola station in Locarno. As all internal observers were assumed to align themselves on the primary observer (Waldmeier during that period), without applying any rescaling by a personal k coefficient, those internal yearly k values can bring invaluable insights on how and to what extent assistants managed to actually align themselves on the primary reference in their daily raw observations. As this internal practice was introduced by Wolf, as soon as 1870, when he started to combine his own counts with those of his first assistants, this can thus help in the understanding of the Zurich number production well before 1945.

\begin{figure}
	\centering
	\includegraphics[trim=25 16 18 21,clip,width=1\linewidth] {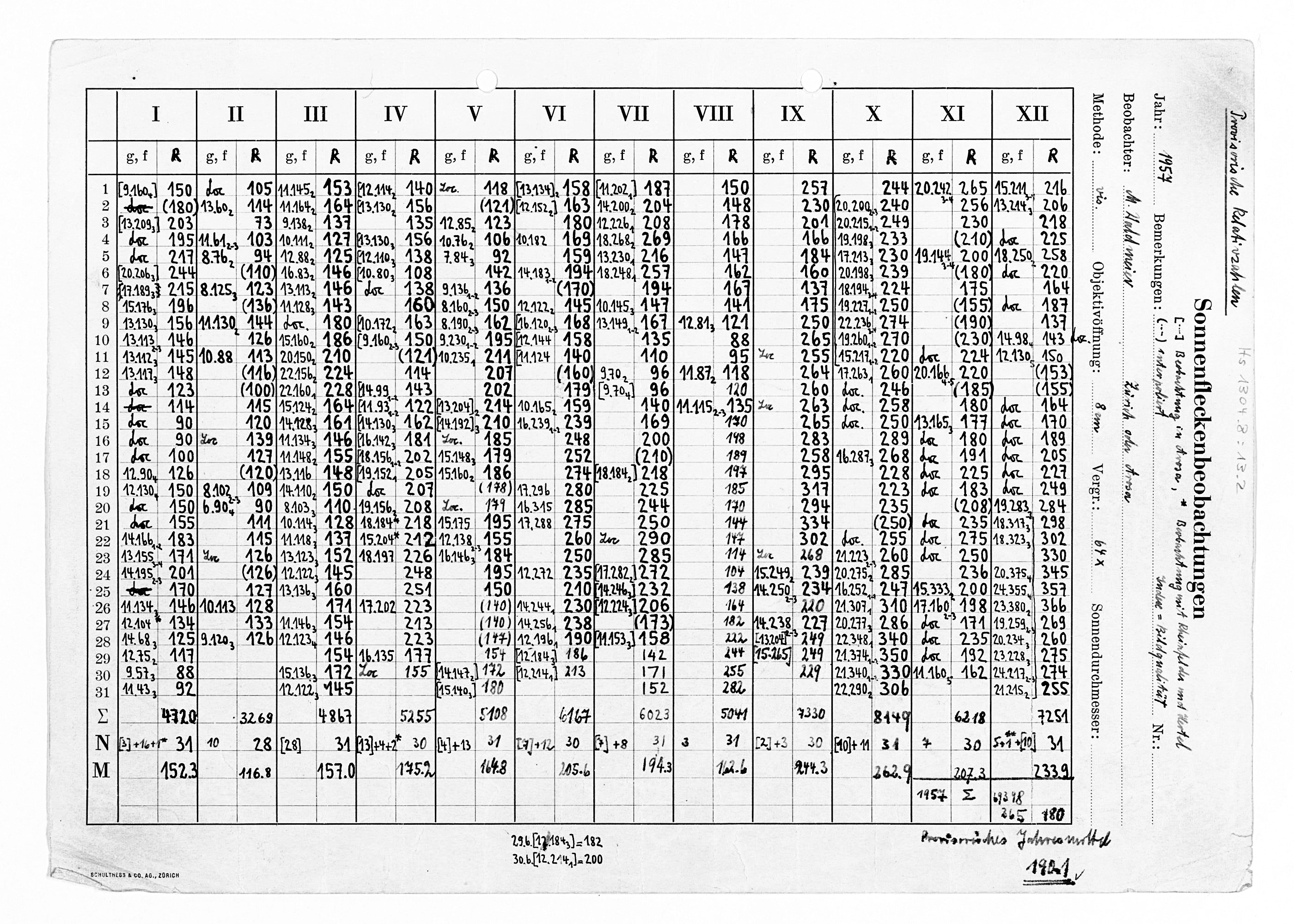}
	\caption{Facsimile of the primary table for Max Waldmeier in 1957, extracted from the complete 1945\,--\,1980 collection of source tables (ETH catalogue entry Hs1304.8:13.2; DOI: {\it 10.7891/e-manuscripta-87246}). Such tables are particularly important, as Waldmeier was the pilot observer of the Zurich sunspot number over that 35-year interval. They include various annotations that allow retracing day-by-day, how Waldmeier himself was observing, and which alternate number was used on days when he could not observe. They thus contain essential information about the Zurich data processing that cannot be found in any other Zurich document.}
	\label{fig:WaldmTable1957}
\end{figure}

In this collection, the most important tables are the yearly tables for the primary observer, Max Waldmeier (Figure \ref{fig:WaldmTable1957}). They provide unique information about three key aspects of the resulting Zurich sunspot numbers. Firstly, those tables were the master tables from which the daily sunspot number was derived for each day of the year. Therefore, they include raw counts and the resulting Wolf number for each day of the year. They thus provide a complete day-by-day census of how each daily SN was derived.

Secondly, most of the days contain the personal counts by Waldmeier, who had the role of base reference. Therefore, this is the yearly table of raw group and sunspot counts by the primary observer, which allows tracking changes in Waldmeier's own daily observations. For instance, Waldmeier was sometimes on mission at the coronagraph of the astronomical station in Arosa, then observing from high altitude with an alternate telescope. The counts for those days may deviate from the base reference scale defined by the standard Fraunhofer refractor used on the front terrace of the observatory in downtown Zurich. Fortunately, Waldmeier marked the days when he observed from Arosa, which will allow analyzing the consequences of this site alternation. 

Thirdly, the days in which Waldmeier could not observe are filled with numbers from local assistants or from the stations in Arosa or Locarno (Karl Rapp until 1 April 1957 and the Specola Observatory starting on 1 October 1957). As can be seen in Figure \ref{fig:WaldmTable1957}, those days are also marked in the tables with a symbol identifying which alternate observer was used. Finally, as those tables record the provisional values issued immediately at the end of each month, on the remaining missing days when none of  the local stations had managed to observe the Sun, the numbers were simply interpolated between adjacent days, and those dates are marked as ``interpolated''. These are the few days which were later replaced by definitive values calculated using k-normalized Wolf numbers from the auxiliary stations, according to a standard method, of which the principle can be reconstructed from a few reference documents \citep{Friedli2020}. 

Those master tables thus provide almost all the keys that were badly missing to reconstruct the method and practices implemented in Zurich, and most probably, to retrace persisting changes or local inconsistencies in the Zurich processing. 

\section{A Major Disruption: Zurich Observers} \label{S-ZuObs}
\begin{figure}
	\centering
	\includegraphics[width=1\linewidth]{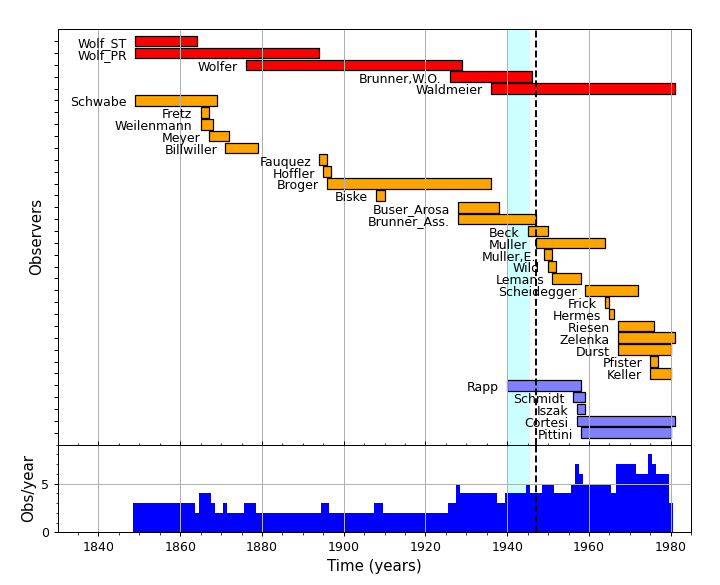}
	\caption{Timelines of the active observing periods of all Zurich observers. In red (top group), the primary observers and in orange (bottom group), the assistants. In purple, the observers of the auxiliary station in Locarno, who were considered as members of the Zurich core group. The vertical shaded band marks World War II and the vertical dashed line indicates the time when the 1947 scale jump occurs in the original SN series. The bottom plot gives the number of active Zurich observers for each year.}
	\label{fig:TimelinesZurich}
\end{figure}

Although the above data still need to be digitized, we now have the full list of observers who contributed year-by-year to the Zurich sunspot number up to 1980. By assembling the timelines of each individual observer, we could map how their observing period overlaps with other observers. Figure \ref{fig:TimelinesZurich} retraces the observing periods for all Zurich primary observers and all the assistants, between 1850 and 1960. 

In this figure, Schwabe is included among the assistants (orange group) although he was an external observer. Indeed, before Wolf could recruit his first assistants in the newly founded Zurich Observatory in 1865, he used Schwabe's numbers as primary alternate source for filling the gaps in his own observations, and even initially considered Schwabe's numbers as fully equivalent to his own (personal k = 1) before 1859. We also included K. Rapp in the associated Locarno station (purple group) although he contributed before the establishment of the Specola Observatory by Waldmeier in 1957, starting in 1940. Indeed, both Brunner and Waldmeier always included Rapps's data together with the Zurich data in the Mitteilungen, even when the data of all the other external stations were not published anymore. Rapp was also trained to follow the Zurich observing methods, and can thus be considered as an internal member of the Zurich group of stations. Finally, although Waldmeier, the last primary observer (red group), started observing as an assistant in 1936, his participation was partly interrupted, as explained below.

For the period before 1944, the resulting chronology reveals a few interesting facts. In particular, one of Wolfer's assistants, Max Broger, had a very long observing career (40 years, 1896 to 1935). He actually observed over more years than several primary observers. As he observed in parallel with Wolfer and then with Brunner, his observations can provide an essential link to check the Wolfer-Brunner homogeneity. 

This touches the fundamental issue of the weighted sunspot counts used by the Zurich Observatory, as mentioned in the previous section. Indeed, \citet{CletteEtal2014}, \citet{CletteLefevre2016}, and \citet{SvalgaardEtal2017} conclude that this alternate counting method is the most likely cause of the 1947 scale jump in the original SN series. However, the timing and sharpness of the jump seem to be contradicted by the fact that this weighting practice was introduced progressively well before 1947, in the early $\rm 20^{th}$ century by Wolfer \citep{CortesiEtal2016, SvalgaardEtal2017}. Although Wolfer himself never used it for his own counts \citep{SvalgaardEtal2017}, this practice was implemented to help assistants aligning their raw counts on the reference of Wolfer, the primary observer. This could be verified by taking the counts on occasional days when only a single big spot was visible on the Sun. Then, when one of Wolfer's assistants, W.O. Brunner, took over as director and as primary observer in 1926, he continued to use weighted counts, but this time as primary observer. Although this marks the moment when the break with Wolf's original methodology occurred, Brunner managed to maintain the stability of his counts, as found by \citet{SvalgaardEtal2017}. When Waldmeier took his succession in 1945, after being assistant for a few years, he thus just continued an established practice. So, apparently, this chronology does not match at all the abrupt occurrence of a jump in 1947, two years after Waldmeier became the new reference observer, a status that he kept for the next 35 years without any other noticeable transition.

Now, by retracing the composition of the network of collaborating observers, we found evidence of a major transition that occurred between 1945 and 1947. The change was twofold. Firstly, at the Zurich Observatory, although Waldmeier became director in 1945, the former director, W.O. Brunner, actually continued observing during one year until December 1945 (see Figure \ref{fig:TimelinesZurich}). Moreover, his primary assistant, W. Brunner-Hagger, who was part of the team since 1928, continued until August 1946. This actually marks the moment when the link with the former Zurich core team is broken. As shown in Figure \ref{fig:TimelinesZurich}, in 1945, Waldmeier starts to recruit new assistants. However, the first one, Beck, worked in parallel with Brunner only during a few months, when solar activity was rather low, and he left the observatory already in 1949. Then follows a succession of other assistants who also leave after only a few years. This means that the overlap between the old and new team was extremely limited and that for several years the Zurich team was very unstable, contrary to the Brunner team that had remained unchanged for nearly 20 years.

So, the internal stability of the Zurich system during the 1945 Brunner-Waldmeier transition rested only on Waldmeier himself. This is unprecedented in the entire Zurich history. Indeed, the stability of the Wolf-Wolfer transition benefited from a 17-year period, during which Wolf and Wolfer observed jointly. Although the Wolfer-Brunner joint period was shorter (3 years, 1926-1928), another assistant, Broger brought a solid reference to bridge the Wolfer-Brunner transition, as he had worked jointly with Wolfer for 30 years, since 1896, and then continued for 10 years together with Brunner, until 1935. 

Finally, although Waldmeier started collaborating with the Zurich Observatory in 1936, he did not contribute during three years, from 1939 to 1941 because of the onset of World War II. Moreover, as he was strongly involved in coronagraph observations, he worked for a large part of his time at the Arosa station, rather than as an ordinary assistant observing side by side with Brunner in Zurich. We also note that the last years before 1946 fell in a minimum of the solar cycle, when the low sunspot activity makes mutual comparisons less accurate. Therefore, all those circumstances reduced the effective overlap period between Brunner and Waldmeier.

\section{A Major Disruption: Auxiliary Stations} \label{S-AuxSta}

\begin{figure}
	\centering
	\includegraphics[width=1\linewidth]{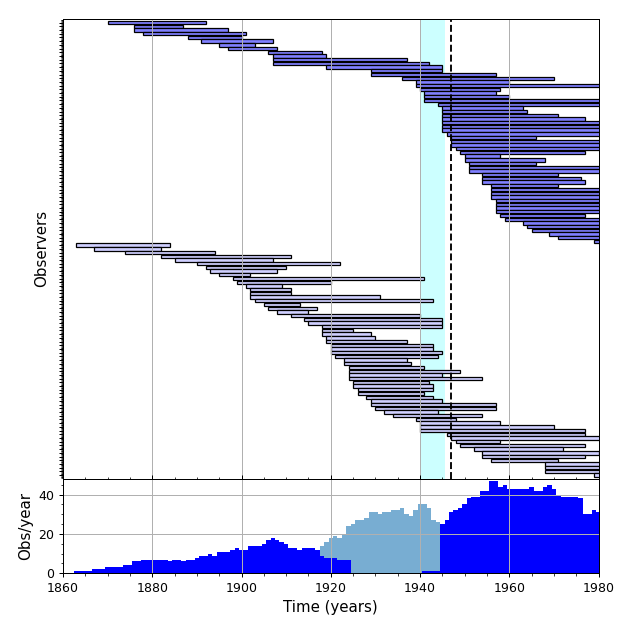}
	\caption{Timelines of the active observing periods of all external stations that sent data to Zurich until the observatory was closed in 1980. The stations are ordered according to the starting date of their series. The top series (dark blue) gathers the professional observatories and the bottom group (light blue) gathers the individual amateur observers. The vertical shaded band marks World War II and the vertical dashed line indicates the 1947 scale jump in the original Zurich series. The bottom plot gives the total number of active stations per year. The light-blue section indicates unpublished data that have not been recovered yet in the Zurich archives.}
	\label{fig:TimelinesStations}
\end{figure}

In parallel with the Zurich internal transition, another major and unprecedented disruption also occurred just after 1945, but now for the Zurich auxiliary stations. Although those external data were not at the core of published sunspot numbers, they definitely provided a wide ensemble of independent data series against which the Zurich numbers were continuously compared. Moreover, all external stations derived their counts using Wolf's original definition, without any weighting. Therefore, the auxiliary data were not affected by the introduction of Zurich's internal weighting practice, and in that sense, they provided the only base against which the Zurich team could infer that their weighted numbers remained coherent with the unwheighted Wolf numbers that formed the original SN series until Wolfer's retirement in 1926 \citep{CletteEtal2014,SvalgaardEtal2017}. This continuous bench-marking could only work if at any given time, there was a large number of active auxiliary stations which had already contributed data during many past years, preferably over one or more full solar cycles.    

Figure \ref{fig:TimelinesStations} shows the timelines of all auxiliary stations that contributed observations to the Zurich Observatory since the mid-$\rm 19^{th}$, over a duration longer than 11 years, i.e. a full solar cycle. This subset of long-duration stations is indeed the most important for the long-term calibration and stability of the series. We distinguished the professional observatories from the individual amateur observers, which reveals a deep evolution in the composition of the Zurich observing network. While a large majority of stations were individual observers before World War II (WWII), professional observatories dominate the network after WWII.

However, a much more drastic change is also caused by WWII. In Figure \ref{fig:TimelinesStations}, we see that, starting in 1938, long-time contributing stations cease to send data, one after the other. When WWII ended, none of those stations, which gave an external benchmark for the earlier Zurich SN, had survived. During the war, given the steep drop of contributing stations, Brunner and Waldmeier called to the rescue a large number of local Swiss amateur astronomers, but this local network was quickly changing, as most observers contributed only for one year or at best a few years (therefore, they do not appear in Figure \ref{fig:TimelinesStations}). None of those observers were long-term observers in the preceding Zurich network established by Wolfer and Brunner. 

Then, just after the war, Waldmeier quickly undertakes the construction of a new international network. The number of stations grows steeply and reaches about 50 stations (see Figure \ref{fig:MittStat}), a number that will remain rather stable until 1980. As noted before, this new network includes many professional observatories, which since then, have delivered observations over very long durations. In fact, some of them are still contributing nowadays to the worldwide SILSO network, and thus provide an invaluable long-term reference spanning up to 75 years, since 1945. However, none of those new stations were part of the pre-1940 long-term network. Therefore, the context in which the sunspot number was produced after 1945 was largely disconnected from the context surrounding this production before 1940. This further weakened the thin internal continuity within the Zurich Observatory. 

\begin{figure}
	\centering
	\includegraphics[width=1\linewidth]{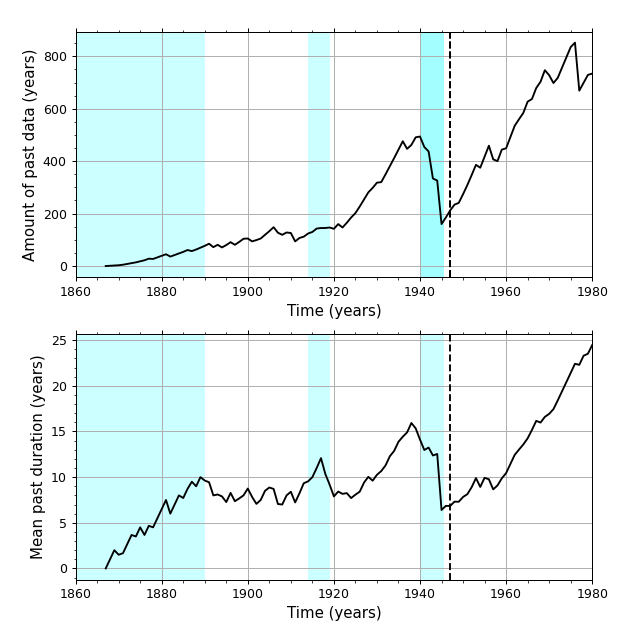}
	\caption{Evolution of the amount of past data available for each year at Zurich. The upper plot gives the total number of preceding observed years by all the stations active on a given year. After an almost continuous increase, a sharp drop occurs just after WWII. The lower plot shows the mean number of preceding observed years per station, for all stations active on a given year. The rise after 1925 indicates the growing participation of stations with very long duration, but a drop to 19$\rm ^{th}$ century levels marks the late 1940's and early 1950's.     }
	\label{fig:NpastData}
\end{figure}

In order to give a more quantitative measure of this second disruption, we summed the number of past observed years already accumulated by all stations that were active on a given year. Figure \ref{fig:NpastData} (top plot) shows the temporal evolution of this total number, which gives a measure of the total amount of past information that the Zurich Observatory had at its disposal for past comparisons and the verification of their stability relative to independent observers.  As expected, the evolution is characterized by a steady increase in the total amount of available data. The only interruption in this trend is the steep drop during WWII, when the count suddenly drops back to the values of the early $\rm 20^{th}$ century. After WWII, there is a recovery, but it takes about 15 years before the amount of past reference data comes back to the value just before WWII. Afterwards, the amount of past data from active stations continues to grow and finally stabilizes in the 1970's.

If we divide this total number of past observed years by the number of active stations, we obtain the mean past duration over which stations active at a given time have been observing before that time (Figure \ref{fig:NpastData}; bottom plot). This mean duration quantifies the past memory built into the SN system. Between 1860 and 1890, this mean duration increases. This marks the progressive recruiting of the first auxiliary observers by Wolf. Then, the mean duration largely stabilizes until 1926, i.e. the Wolfer-Brunner transition. The only feature is a temporary peak associated with WWI, which thus left only a minor imprint in this evolution. Thanks to the many new observers recruited after WWI, and who continue observing until WWII, the mean duration grows to almost 15 years in 1938. 

Then, WWII again produces a steep drop, by a factor of two. In 1945 and the decade that follows, the mean memory range falls back to about 7 years, a level that was not encountered since 1880, i.e. the epoch when Wolf was still recruiting his first associated observers. After this dramatic shortening of the past memory, there was a steady recovery. However, it is only around 1965, 20 years after WWII, that the pre-WWII mean memory range is recovered. It continues to rise until 1980, when the Zurich Observatory was closed. This continuous trend largely rests on the long-term contribution from the professional observatories that entered the network just after WWII. 
Figure \ref{fig:NpastData} thus illustrates that the years immediately following WWII were abruptly affected by a major loss of past references, and that this loss had no equivalent in the history of the Zurich SN number. 

Although the above indicators are indirect contextual elements, the fact that this unique double discontinuity in the history of the Zurich sunspot number production coincides with the jump revealed by the SN series itself is a very strong indication that the sharp SN scale jump was a consequence of this abrupt and radical change in the base data input. Until 1946, the potential biasing effect, which was present since the weighted counting method had been introduced, had been kept under control thanks to the double stabilizing effect of long-term internal and external observers who did not change their counting practices. This stabilizing continuity was clearly broken between 1946 and 1947, which suddenly opened the way for the biasing effect inherent to the weighted counts, as evidenced by the 1947 upward jump. This new contextual evidence thus explains simultaneously the delayed effect of the weighting practice and the abruptness of the jump.

\section{Conclusion} \label{S-Conclu}

Over just a few years, we thus achieved major progress in the construction of the SN database. Now, about two thirds of the existing source data are recorded in digital form. We can now also report on the recovery of a major missing part of this collection, the yearly source tables of the Waldmeier era from 1945 to 1980. This fills the main gap in the SN database and provides the missing link between the contemporary index and the rest of this long series before 1945 and back to 1700.
While significant work is still needed to digitize those newly recovered documents, the global panorama that the SN database now offers made it possible to establish the complete chronology of contributing stations and observers. We found that the two world wars had deep consequences on the production of the SN by the Zurich Observatory. WWI brought a major expansion of the network of auxiliary observers, but without disrupting the internal practices and organization of the Zurich sunspot observers. 

On the other hand, after WWII, we find a double disruption in the Zurich system. A complete renewal of the Zurich observing team occurred between 1946 and 1947, with almost no overlap between the old team, which had remained mostly unchanged for more than 20 years, and the new team progressively built by Waldmeier between 1946 and 1950. Moreover, after the loss of most of the external observers active over the decades preceding WWII, between 1938 and 1945, an entirely new worldwide network is established after the war with entirely different stations. The narrow correspondence of this drastic and unprecedented structural change with the 18\% SN scale-jump diagnosed in the SN series provides strong historical evidence that a sharp jump in the SN exactly at that moment is a real and logical consequence. Although the suspected cause, i.e. the introduction of the size-based weighting of the spot counts, was introduced much earlier in the practice of Zurich assistants, our now-complete timeline explains why it only led to actual consequences when this sharp and unprecedented discontinuity in the Zurich system took place. 

All together, those recovered tables open the way to future major steps in the end-to-end calibration of the sunspot number series.  Full statistical diagnostics of the actual stability of each separate Zurich observer, which was simply postulated since the epoch of Wolf, will allow disentangling in detail the causes of anomalies found in the heritage series. Much more importantly, those data open the way for a full recalculation of the sunspot number, starting again from the full set of raw input data. This recalculation will use new advanced computer-based processing methods, which exploit the entire set of data instead of mostly using the numbers of the single primary observer, as was the case in the original Zurich series. This should improve further the stability and accuracy of the sunspot number in the interval 1945-1980, where so far, SN Version 2 consisted only in a correction factor applied to the original Zurich SN series. This would also finally bridge the gap separating the current international sunspot number from the early epoch before 1945.

However, a partial gap still remains. Although all observations made in Zurich from Wolf in 1849 to Waldmeier in 1980 now finally form a complete and uninterrupted thread, we still miss the unpublished archives from the Brunner era. Therefore, efforts are still continuing to try recovering the last missing data from the network of the auxiliary stations between 1919 and 1944. Hopefully, this will finally bring the last touch to this digital database that will feed sunspot science and long-term solar-cycle studies for many years.   


\begin{acks}
This work and the team of the World Data Center SILSO (\url{http://www.sidc.be/silso/}), which produces the international sunspot number and maintains the sunspot database used in this study, are supported by Belgian Solar-Terrestrial Center of Excellence (STCE, \url{ http://www.stce.be}) funded by the Belgian Science Policy Office (BelSPo). This work was also supported by the International Space Science Institute (ISSI, Bern, Switzerland) via the International Team 417 ``Recalibration of the Sunspot Number Series'', chaired by M. Owens and F. Clette (\url{https://www.issibern.ch/teams/sunspotnoser/}). Specola Solare Ticinese acknowledges the financial support provided by Canton Ticino through the Swisslos fund and by the Federal Office of Meteorology and Climatology MeteoSwiss, in the framework of GCOS. We would like to thank Thomas Friedli for digitizing and making available the original sourcebook by R. Wolf via the web site of the Rudolf Wolf Society (\url{http://www.wolfinstitute.ch}). We also thank the ETH Library (\url{https://library.ethz.ch/en/}), and in particular Evelyn Boesch, of the Hochschularchiv, for the deep searches in the catalogues and archives, and for giving us access to original historical documents from the Zurich Observatory. We also thank Olivier Lema\^itre for developing the software and computer database, Stephen Fay and Shreya Bhattasharya for the quality control, and last but not least, we are also grateful to the summer-job students who patiently and carefully encoded all numbers tabulated in the original paper documents: Elfaniel Hermel, Esther-Lauren M'Bilo and Mael Panouillot.
\end{acks}

\vspace{\baselineskip}

\textbf{Disclosure of Potential Conflicts of Interest}\\
The authors declare that they have no conflicts of interest.




\end{article} 
\end{document}